\documentclass[aps,twocolumn,amsmath,amssymb,floatfix,superscriptaddress]{revtex4-1}
\usepackage{tabularx}
\usepackage{bm}
\usepackage{epsfig,psfrag,subfigure}
\usepackage{graphicx}
\usepackage{color}
\usepackage{amsfonts}
\usepackage{exscale}
\usepackage{amsbsy}
\usepackage{multirow}

\usepackage{times}
\usepackage{float}
\usepackage{latexsym,amsmath,amssymb,bm,euscript}
\usepackage{epstopdf}
\usepackage[colorlinks=true,linkcolor=blue,citecolor=blue]{hyperref}
\usepackage{soul}
\usepackage[normalem]{ulem}
\usepackage{mathrsfs}
\usepackage{amsmath}
\usepackage{lettrine}
\usepackage{bbding}
\usepackage{xspace}
\usepackage{textcomp}
\usepackage{textcase}
\usepackage{setspace}

\usepackage[dvipsnames]{xcolor}
\def\avg#1{\left\langle#1\right\rangle}

\def\trace#1{{\rm Tr}\left[#1\right]}

\def\be{\begin{equation}}       \def\ee{\end{equation}}
\def\bea{\begin{eqnarray}}      \def\eea{\end{eqnarray}}
\def\ba{\begin{array} }
\def\ea{\end{array} }

\def\nn{\nonumber}

\def\=>{\Rightarrow}
\def\>{\rightarrow}

\def\Fig#1{Fig.~\ref{#1}}

\renewcommand{\v}[1]{{\bf #1}}

\renewcommand{\>}{\rangle}

\begin{document}
%\title{Parallel DMRG study of spin-$1/2$ triangular Heisenberg model}
\title{Nature of quantum spin liquids of the S=$1/2$ Heisenberg antiferromagnet on the triangular lattice: A parallel DMRG study}
\author{Yi-Fan Jiang}
\affiliation{School of Physical Science and Technology, ShanghaiTech University, Shanghai 201210, China}
\affiliation{Stanford Institute for Materials and Energy Sciences, SLAC National Accelerator Laboratory and Stanford University, Menlo Park, CA 94025, USA}
\author{Hong-Chen Jiang}
\email{hcjiang@stanford.edu}
\affiliation{Stanford Institute for Materials and Energy Sciences, SLAC National Accelerator Laboratory and Stanford University, Menlo Park, CA 94025, USA}

\begin{abstract}
We study the ground-state properties of the quantum spin liquid (QSL) phases of the spin-$1/2$ antiferromagnetic Heisenberg model on the triangular lattice with nearest- ($J_1$), next-nearest- ($J_2$), and third-neighbor ($J_3$) interactions by using density-matrix renormalization group (DMRG) method. By combining parallel DMRG with $SU(2)$ spin rotational symmetry, we are able to obtain accurate results on large cylinders with length up to $L_x=48$ and circumference $L_y=6 - 12$. Our results suggest that the QSL phase of the $J_1$-$J_2$ Heisenberg model is gapped %consistent with a gapped spin liquid 
which is characterized by the absence of gapless mode, short-range spin-spin and dimer-dimer correlations. In the presence of $J_3$ interaction, we find that a new critical QSL with a single gapless mode emerges. While both spin-spin and scalar chiral-chiral correlations are short-ranged, dimer-dimer correlations are quasi-long-ranged which decays as a power-law at long distances.%Moreover, our results suggest that the critical QSL state preserves the time-reversal symmetry without long-range scalar chiral order.
\end{abstract}
\maketitle

%==Introduction==
Quantum spin liquids (QSLs) are highly entangled phases of matter that exhibit novel features associated with their topological character and support factional excitations, yet resist symmetry breaking even down to zero temperature due to strong quantum fluctuations and geometric frustrations.\cite{Balents2010,Savary2016,Broholm2019}
Broad interest in QSLs was triggered by its important role in understanding strongly correlated materials especially high temperature superconductors as well as its potential application in topological quantum computation.\cite{Broholm2019,Anderson1987,Emery1987,Lee2006,Nayak2008,Fradkin2015} One of the most promising systems to realize QSLs is the spin-1/2 Heisenberg antiferromagnet on the triangular lattice which is defined by the model Hamiltonian%
\begin{eqnarray}
H=\sum_{ij} J_{ij} \mathbf{S}_i \cdot \mathbf{S}_j.
\label{Eq:Ham}
\end{eqnarray}
A number of studies of the $J_1$-$J_2$ model with first- ($J_1$) and second-neighbor ($J_2$) exchange couplings have led to a consensus that there is an intermediate QSL phase (referred to as $J_1$-$J_2$ spin liquid) in the range of $0.07 < J_2/J_1 < 0.15$, which is sandwiched by the $120^\circ$ magnetic phase and a stripe magnetic phase.\cite{Jolicoeur1990,Manuel1999,Mishmash2013,Iqbal2016,Zhu2015,Saadatmand2016,Hu2015,Kaneko2014,Hu2019,Li2015,Zheng2015,Hu2016,Gong2017,Bauer2017,Wietek2017,Ferrari2019,Gong2019} However, its precise nature remains still under intense debate where distinct types of QSLs have been proposed including the gapped spin liquid\cite{Hu2015,Zhu2015,Saadatmand2016}, the gapless $U(1)$ Dirac spin liquid\cite{Kaneko2014,Hu2019} and the spin liquid with spinon Fermi surface.\cite{Gong2019} The gapped spin liquid is characterized by a fully gapped excitation spectrum and all the correlations, including the spin-spin, dimer-dimer and scalar chiral-chiral correlations, are short-ranged. While the spin-spin correlation is quasi-long-ranged in both the Dirac and spinon Fermi surface spin liquids, the former is gapless only at specific discrete momenta in the reciprocal space, the latter is gapless in the whole spinon Fermi surface. As a result, a further unbiased study is required to identify the precise nature of the $J_1$-$J_2$ spin liquid phase.

Aside from the $J_1$ and $J_2$ interactions, an additional third-neighbor $J_3$ interaction (referred to as $J_1$-$J_2$-$J_3$ model) has also been considered in recent studies, which was proposed as an important ingredient to understand various magnetic properties of the triangular lattice materials CeFeO$_2$ and CuCrO$_2$.\cite{Kadowaki1990,Kimura2006,Ye2007,Seki2008} Interestingly, recent study\cite{Gong2019} has provided numerical evidences that a new type of chiral spin liquid (CSL) state could be realized in the $J_1$-$J_2$-$J_3$ model, which spontaneously breaks the time-reversal symmetry (TRS) and has long-range scalar chiral order. Distinct with the Kalmeyer-Laughlin state\cite{Kalmeyer1987}, this CSL has a spinon Fermi surface with gapless excitation spectrum. However, the spin-spin correlations decay exponentially which seems inconsistent with the presence of the spinon Fermi surface. To resolve the discrepancy and understand the QSL phase of the $J_1$-$J_2$-$J_3$ model, further numerical simulation is required.

\begin{figure}
    \centering
    \includegraphics[width=\linewidth]{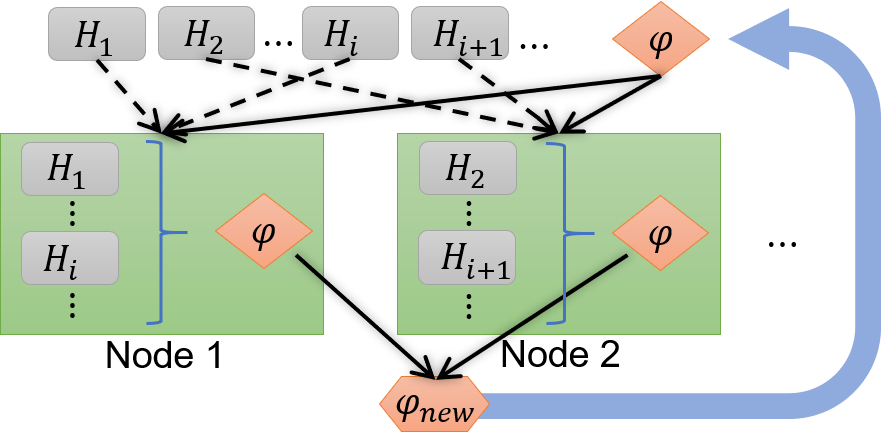}
    \caption{The schematic Lanczos step in the parallel DMRG simulation using two nodes. Each term $H_i$ in the Hamiltonian $H=\sum_i H_i$ is equally distributed to different nodes at the beginning of the Lanczos step. In each iteration, the wave function $\psi$ is copied to different nodes, multiplied by the terms stored in each node and accumulated to form $\psi_{new}$ which is used as the input wave-function for the next iteration.}\label{Fig:pdmrg}
\end{figure}

%==Principal results==
%\textbf{Principal results: }%
In this paper, we address the above questions by studying both the $J_1$-$J_2$ and $J_1$-$J_2$-$J_3$ models on triangular cylinders with circumference $L_y=6 - 12$ and length up to $L_x=48$ using density-matrix renormalization group (DMRG) encoded with $SU(2)$ spin rotational symmetry.\cite{White1992,McCulloch2002,Chan2004} Specifically, we have developed an efficient %operator-level
parallel DMRG scheme and performed both real and complex-value DMRG simulations. The parallel scheme\cite{Chan2004}, which is based on equally distributing the Hamiltonian as illustrated in Fig.\ref{Fig:pdmrg}, has further improved the numerical efficiency by $O(L_y)$ times, so that we are able to keep up to $m=9000$ $SU(2)$ states (equivalent $m=36000$ $U(1)$ states) in the complex-value DMRG simulation to obtain accurate results.

For more reliable results, we focus on typical sets of parameters deep inside the QSL phases of both models used in previous studies.\cite{Hu2015,Zhu2015,Saadatmand2016,Kaneko2014,Hu2019,Gong2019} Our results suggest that the $J_1$-$J_2$ spin liquid is consistent with a gapped QSL,\cite{Moessner2001,Yao2012} where all correlations, including the spin-spin, dimer-dimer and scalar chiral-chiral correlations, are short-ranged which decay exponentially at long distances. In the presence of $J_3$ interaction, we find that a new type of QSL, dubbed critical spin liquid,\cite{Rokhsar1988,Yao2011} emerges in the $J_1$-$J_2$-$J_3$ model. There is a single gapless mode which is independent of the circumference of the cylinders. While both spin-spin and scalar chiral-chiral correlations are short-ranged, the dimer-dimer correlations are quasi-long-ranged.

%==Fig.1 lattice==

\begin{figure}
    \centering
    \includegraphics[width=\linewidth]{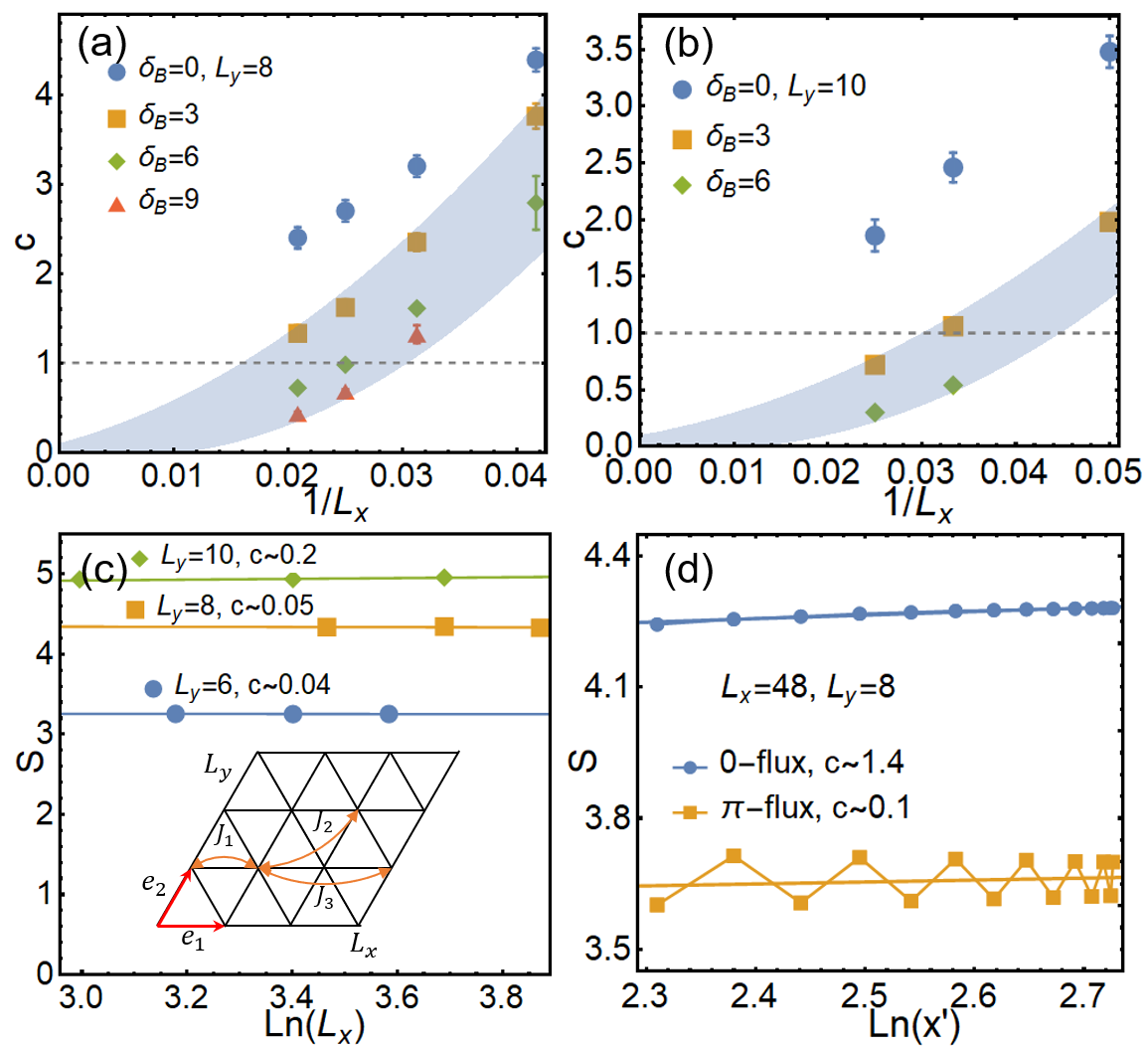}
    \caption{Entanglement entropy $S$ and central charge $c$ for the $J_1$-$J_2$ model. The extracted $c$ with $J_2=0.11$ on (a) $L_y=8$ and (b) $L_y=10$ cylinders using Eq.(\ref{Eq:EE}), where $\delta_B$ is the number of data points omitted from the open boundaries. The shaded region is a guide for eyes. (c) $S(L_x/2)$ as a function of ${\rm Ln}(L_x)$ on $L_y=6 - 10$ cylinders where solid lines denote the fitting $S(L_x/2)\sim \frac{c}{6}{\rm Ln}(L_x)$. (d) $S(x)$ on $L_y=8$ cylinder of length $L_x=48$ with $0$ and $\pi$ flux inserted through cylinder, where $x'=\frac{L_x}{\pi} \sin \frac{\pi x}{L_x}$.}\label{fig:entropyJ2}
\end{figure}

%==Model and Method==
\textbf{Model and Method: }%
We employ DMRG\cite{White1992,McCulloch2002,Chan2004} to study the ground state properties of the spin-1/2 antiferromagnetic Heisenberg model on the triangular lattice defined in Eq.(\ref{Eq:Ham}). The lattice geometry used in our simulations is depicted in the inset of Fig.\ref{fig:entropyJ2}(c), with open (periodic) boundary condition along the $\mathbf{e}_1$ ($\mathbf{e}_2$) direction, where $\v{e}_1=(1,0)$ and $\v{e}_2=(1/2,\sqrt{3}/2)$ are two basis vectors. We focus on cylinders with circumference $L_y$ and length $L_x$, where $L_y$ and $L_x$ are the number of sites in the $\mathbf{e}_2$ and $\mathbf{e}_1$ directions, respectively. We set $J_1=1$ as an energy unit and focus on two typical sets of parameters used in previous studies.\cite{Gong2019} These correspond to the $J_1$-$J_2$ model with $J_2=0.11$, and the $J_1$-$J_2$-$J_3$ model with $J_2=0.3$ and $J_3=0.15$, respectively. In this paper, we report results on $L_y=6 - 12$ cylinders of length up to $L_x= 48$.

We perform both real- and complex-value DMRG simulations and keep up to $m=9000$ $SU(2)$ states (equivalent $m=36000$ $U(1)$ states) in each DMRG block. % to obtain accurate results. 
To conquer the extensive numerical cost, especially, on wide cylinders with large number of states, we have developed an efficient operator-level parallel DMRG scheme with $SU(2)$ spin rotational symmetry. The Hamiltonian on a width $L_y$ cylinder in the block-site-site-block decomposition\cite{White1992} typically has $\sim \alpha L_y$ terms, where the coefficient $\alpha$ depends on the number of independent operators (e.g. $\{ S^+, S^z\}$), the coordination number of the lattice and the specific form of interactions in the Hamiltonian. For instance, $\alpha\sim 2$ in the $SU(2)$ DMRG simulation, while $\alpha\sim 6$ in the $U(1)$ DMRG simulation for both $J_1$-$J_2$ and $J_1$-$J_2$-$J_3$ models.

As sketched in Fig.\ref{Fig:pdmrg}, in each step of the DMRG simulation, we equally distribute the decomposed $\alpha L_y$ terms to $n$ nodes, based on which the most time-consuming Lanczos part can be accelerated by $n$ times (up to $\alpha L_y$ times). Other parts of the DMRG simulations can be parallelized similarly. An obvious advantage of the operator-level parallel scheme over the real-space parallel scheme\cite{Stoudenmire2013} is that it is does not introduce any additional approximation compared with the single-node DMRG scheme. Besides the operator-level parallelism, parallel DMRG simulation can also be achieved by distributing matrix-vector contraction or blocks with different quantum numbers to the working nodes or generalizing the two-sites DMRG algorithm to the N-sites version.\cite{Julian2010, Yamada2011, Kantian2019, Levy2020} Further details of the operator-level parallel DMRG scheme are provided in the Supplemental Material (SM).

%==Fig.2 J1-J2 entropy==

{\bf $J_1$-$J_2$ model: }%
The central debate on the $J_1$-$J_2$ spin liquid is whether it is gapped or gapless. A key diagnostic to distinguish distinct types of QSLs proposed in the previous studies is the number of gapless spin modes, i.e., the central charge $c$. The gapped QSL has no gapless spin mode with $c=0$.\cite{Hu2015,Zhu2015,Saadatmand2016} For the $U(1)$ Dirac spin liquid, $c\leq 3$ which depends on the momentum cut across the Dirac points.\cite{Kaneko2014,Hu2019} On the contrary, for the spin liquid with spinon Fermi surface, the value of $c$ increases with the width $L_y$ of the systems.\cite{Gong2019} %, e.g., $c=1$ on the $L_y=6$ cylinders and $c=5$ on the $L_y=8$ cylinders.\cite{Gong2019}

 %==Fig.3 J1-J2 spin and dimer correlation==
\begin{figure}
    \centering
    \includegraphics[width=\linewidth]{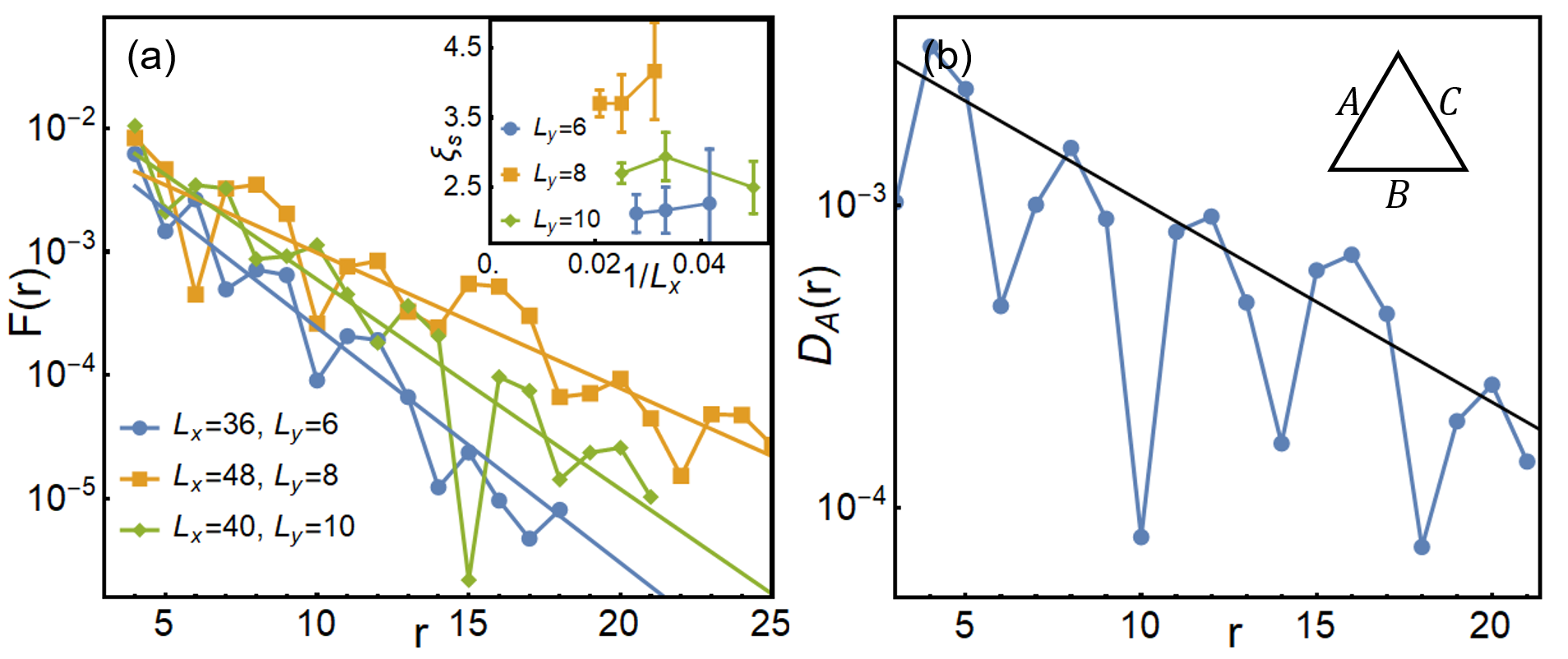}
    \caption{Spin-spin $F(r)$ and dimer-dimer $D_A(r)$ correlations of the $J_1$-$J_2$ model. (a) $F(r)$ with $J_2=0.11$ on $L_y=6 - 10$ cylinders, where solid lines denote the exponential fitting $F(r)\sim e^{-r/\xi_s}$. Inset: Correlation length $\xi_s$ on $L_y=6 - 10$ cylinders as a function of $1/L_x$. (b) $D_A(r)$ on $L_y=8$ cylinder of length $L_x=40$, where solid line denotes exponential fitting $D_A(r)\sim e^{-r/\xi_A}$. $A$, $B$ and $C$ denote the three different bonds.}\label{fig:spincorJ2}
\end{figure}

To better identify the nature of the $J_1$-$J_2$ spin liquid, we focus on $J_2=0.11$ which is deep inside the spin liquid phase of the $J_1$-$J_2$ model. We first calculate the von Neumann entanglement entropy $S(x)=-\trace{\rho_x \ln \rho_x}$ on numerous cylinders where $\rho_x$ is the reduced density matrix of the subsystem with length $x$. For critical system of length $L_x$ with open boundaries, it has been established that $c$ can be obtained using\cite{Calabrese2004, Fagotti2011}%
\bea
S(x)&=&\frac{c}{6} \ln \big[\frac{L_x}{\pi} \sin \frac{\pi x}{L_x}\big]+ const, \label{Eq:EE}
\eea
where examples are shown in Fig.\ref{fig:entropyJ2}. It should be noted that notable finite-size and boundary effects have been observed associated with Eq.(\ref{Eq:EE}), from which $c$ could be dramatically overestimated. To extract $c$ more reliably, we have systematically analyzed both the boundary and finite-size effects. Specifically, for a given cylinder of length $L_x$, we extract $c$ by removing $\delta_B$ data points from both open ends (see SM for details). As shown in Fig.\ref{fig:entropyJ2}(a-b), the extracted $c$ decreases monotonically and rapidly with the increase of both $L_x$ and $\delta_B$. It is worth mentioning that for a given cylinder of length $L_x$, reduced boundary effect by removing several data points from the open ends can provide more reliable results that are much closer to that in the long cylinder limit. In the long cylinder limit $L_x\rightarrow \infty$, i.e., $1/L_x\rightarrow 0$, we find that $c\sim 0$ for $L_y=6 - 10$ cylinders. This suggests that the $J_1$-$J_2$ spin liquid is gapped without gapless spin mode.

As a further test, we have also studied the effect of twisted boundary condition, for instance, anti-periodic boundary condition by inserting $\pi$-flux through the cylinder. Fig.\ref{fig:entropyJ2}(d) shows an example of $S(x)$ on $L_y=8$ cylinder of length $L_x=48$ with periodic (0-flux) and anti-periodic ($\pi$-flux) boundary conditions. The extracted central charge with $\pi$-flux is $c\sim 0.1$, which is much closer to $c=0$ than the normal cylinder. Alternatively, $c$ can be obtained using $S(L_x/2)=\frac{c}{6}\ln (L_x)+const$ as shown in Fig.\ref{fig:entropyJ2}(c), which is $c=0.10(1)$ and $c=0.09(5)$ for $L_y=8$ and $L_y=10$ cylinders, respectively. Similar behavior has also been observed on $L_y=12$ cylinders (see SM for details). All of these are consistent with a gapped state without gapless spin mode.

The absence of gapless mode suggests that all correlations are short-ranged. To see this, we first calculate the spin-spin correlation function defined as%
\begin{eqnarray}
F(r)=|\avg{\v{S}_{(x_0,y_0)}\cdot \v{S}_{(x_0+r,y_0)}}|.\label{Eq:SpinCor}
\end{eqnarray}
Here $\mathbf{S}_{(x_0,y_0)}$ is the spin operator on the reference point $(x_0,y_0)=(L_x/4,L_y/2)$ and and $r$ is the distance between two sites in the $\mathbf{e}_1$ direction.. Fig.\ref{fig:spincorJ2}(a) shows examples of $F(r)$ for $L_y=6 \sim 10$ cylinders. For all cases, $F(r)$ decays exponentially at long distances and can be well fitted by an exponential function $F(r)\sim e^{-r/\xi_s}$ with finite correlation length $\xi_s$ shown in the inset of Fig.\ref{fig:spincorJ2}(a). The fact that $\xi_s$ decreases with the increase of $L_y$ when $L_y\geq 8$ (see SM for details) suggests a finite $\xi_s$ in two dimensions.

We have also measured the dimer-dimer correlation function defined as 
\bea
D_a(r)&=&\left\langle(\hat{B}_a(x,y)-\langle \hat{B}_a(x,y) \rangle ) \right. \cdot \nn\\
& & \left. (\hat{B}_a(x+r,y)-\langle\hat{B}_a(x+r,y)\rangle) \right\rangle.\label{Eq:Dimer}
\eea
Here $\hat{B}_a(x,y)= \v{S}(x,y)\cdot \v{S}(x_a,y_a)$ is the dimer operator on bond type $a=A/B/C$ shown in Fig.\ref{fig:spincorJ2}(b). We find that while the strength of $B_a=\avg{\hat{B}_a(x,y)}$ depends on $a$ due to the broken $C_3$ rotational symmetry of the cylindrical geometry, it has no any spatial oscillation in the bulk of the systems, suggesting the absence of static long-range dimer order. This is further evidenced by the fact that $D_a(r)$ decays exponentially as $D_a(r) \sim e^{-r/\xi_a}$ as shown in Fig.\ref{fig:spincorJ2}(b) with finite correlation length $\xi_a$, e.g., $\xi_{a}\sim 6.5$ on $L_y=8$ cylinder.

%==Fig.4 J1-J2-J3 entropy==
\begin{figure}
    \centering
    \includegraphics[width=\linewidth]{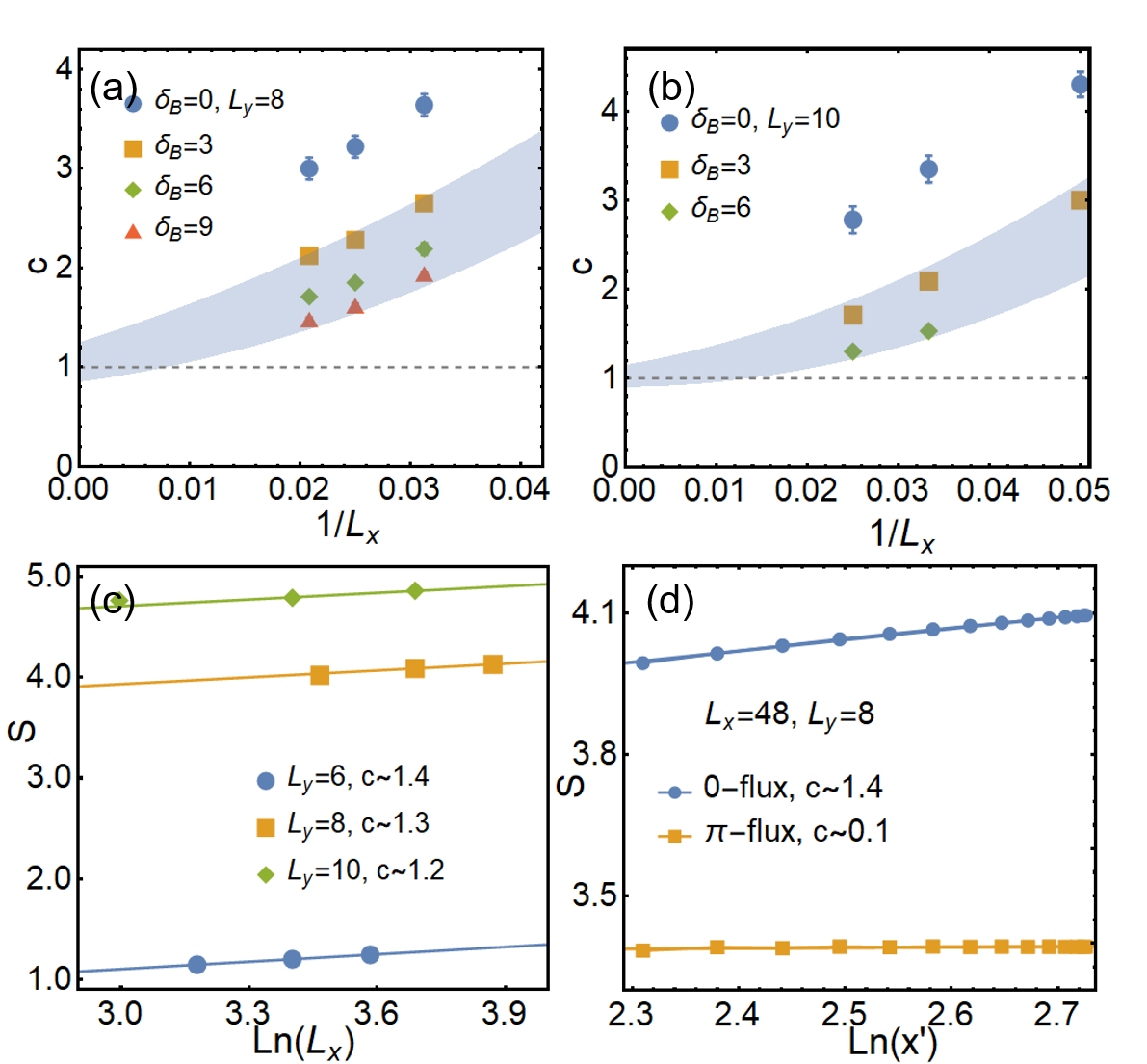}
    \caption{Entanglement entropy $S$ and central charge $c$ for the $J_1$-$J_2$-$J_3$ model. (a) The extracted $c$ with $J_2=0.3$ and $J_3=0.15$ on (a) $L_y=8$ and (b) $L_y=10$ cylinders, where $\delta_B$ is the number of data points omitted from the open boundaries. The shaded region is a guide for eyes. (c) $S(L_x/2)$ as a function of ${\rm Ln}(L_x)$ where solid lines denote the fitting $S(L_x/2)\sim \frac{c}{6}{\rm Ln}(L_x)$. (d) $S(x)$ on $L_y=8$ cylinder of length $L_x=48$ with $0$ and $\pi$ flux inserted through the cylinder. The solid lines denote the fitting $S(x)\sim \frac{c}{6}{\rm Ln}(x')$ where $x'=\frac{L_x}{\pi} \sin (\frac{\pi x}{L_x})$.}\label{fig:entropyJ3}
\end{figure}

{\bf $J_1$-$J_2$-$J_3$ model: }%
In the presence of $J_3$ interaction, recent study\cite{Gong2019} suggests that a distinct QSL state, i.e., a gapless CSL with spinon Fermi surface, can be realized in the $J_1$-$J_2$-$J_3$ model. To rule out the possible finite-size effect, we follow the same procedure with the $J_1$-$J_2$ model. For simplicity, we focus on the same set of parameter as Ref.\cite{Gong2019}, i.e., $J_2=0.3$ and $J_3=0.15$, which is deep inside the QSL phase. We first benchmark our calculations using the same parameters and have observed consistency for both $L_y=6$ cylinders and $N=16\times 8$ cylinder.\cite{Gong2019} (See SM for details.) However, similar with the $J_1$-$J_2$ model, we find that the extracted $c$ on $L_y=8$ cylinders suffers from notable finite-size and boundary effects, which decreases monotonically with the increase of $L_x$ as shown in Fig.\ref{fig:entropyJ3}(a). In the long cylinder limit $L_x\rightarrow \infty$, it approaches to a much smaller value $c\sim 1$, suggesting that there is only one gapless mode on $L_y=8$ cylinder. This is also true on $L_y=10$ cylinders where we also find $c\sim 1$ as shown in Fig.\ref{fig:entropyJ3}(b). It is worth noting that in the limit $L_x=\infty$, our results show that $c\sim 1$ on all $L_y=6-12$ cylinders (see SM for details) without notable dependence on $L_y$, suggesting that there is one gapless mode in the bulk of the system in two dimensions. It is hence reasonable to expect that the single gapless mode may carry momentum $k_2=0$ which is shared by all cylinders. To support this, we have further calculated $S(x)$, e.g., on $N=48\times 8$ cylinder, by inserting a $\pi$ flux through the cylinder where the momentum $k_2=0$ is unavailable. As expected, we find that $c\sim 0.1$ (see Fig.\ref{fig:entropyJ3}(d) inset) which is consistent with the absence of gapless mode.

%==Fig.5 J1-J2-J3 spin and dimer correlation==
\begin{figure}
    \centering
    \includegraphics[width=\linewidth]{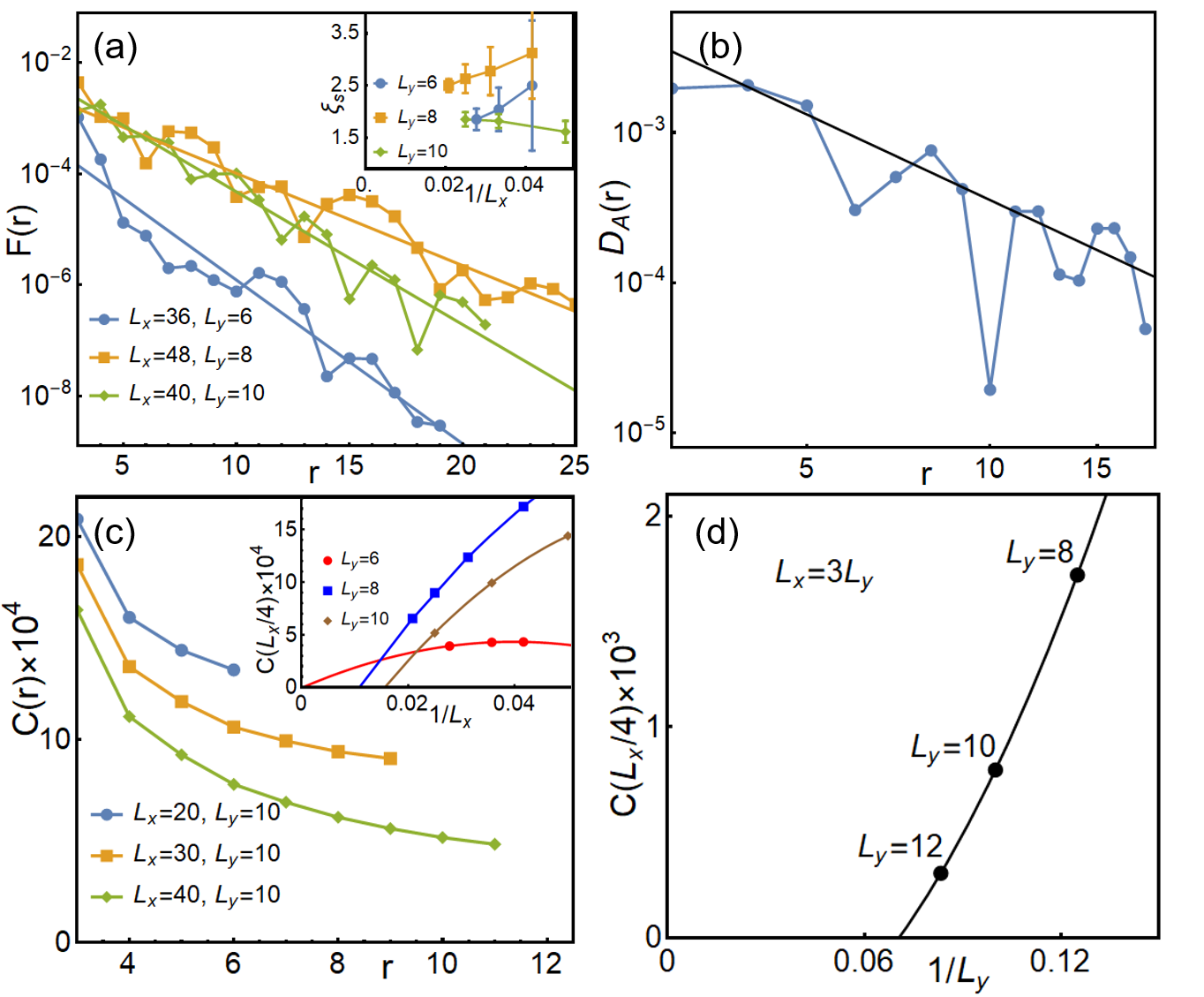}
    \caption{Correlation functions for the $J_1$-$J_2$-$J_3$ model. (a) Spin-spin correlation $F(r)$ with $J_2=0.3$ and $J_3=0.15$ on $L_y=6\sim 10$ cylinders. Solid lines denote the exponential fitting $F(r)\sim e^{-r/\xi_s}$. Inset: correlation length $\xi_s$ as a function of $1/L_x$. (b) Dimer-dimer correlation $D_A(r)$ on $L_y=8$ cylinder of length $L_x=40$, where solid line denotes the power-law fitting $D_A(r)\sim r^{-K_A}$. (c) Scalar chiral-chiral correlation $C(r)$ on $L_y=10$ cylinders. Inset: Finite-size scaling of $C(L_x/4)$ on $L_y=6 - 10$ cylinders as a function of $1/L_x$ using the second-order polynomial function. (d) Finite-size scaling of $C(L_x/4)$ as a function of $1/L_y$ on lattices with fixed ratio $L_x/L_y=3$ using the second-order polynomial function.}\label{fig:spincorJ3}
\end{figure}

We have also calculated the spin-spin correlation $F(r)$ as shown in Fig.\ref{fig:spincorJ3}(a) for $L_y=6 - 10$ cylinders. For all cases, we find that $F(r)$ is short-ranged which decays exponentially at long distances as $F(r)\sim e^{-r/\xi_s}$. Similar with the $J_1$-$J_2$ model, the correlation length is finite $\xi_s=1.5 - 3$ as shown in the inset of Fig.\ref{fig:spincorJ3}(a). (see SM for more details) Contrary to the spin-spin correlation, we find that the dimer-dimer correlation decays as a power-law at long distances as $D_a(r)\sim r^{-K_a}$ with a finite exponent $K_a$, for instance, $K_a\sim 1.8$ on $L_y=8$ cylinder as shown in Fig.\ref{fig:spincorJ3}(b). It is hence reasonable to conclude that the quasi-long-range dimer-dimer correlation is responsible for the single gapless mode.

To test the possibility of TRS breaking, we have measured the scalar chiral-chiral correlation function defined as
\begin{equation}
C(r)=\langle \hat{\chi}_{i_0} \hat{\chi}_{i_0+r}\rangle.\label{Eq:Chiral}
\end{equation}
Here $\hat{\chi}_i=\v{S}_i\cdot( \v{S}_j \times \v{S}_k)$ is the scalar chiral operator defined on a small triangle, $i_0=(x_0,y)$ is the reference point with $x_0=L_x/4$ and $r$ is the distance between two triangles in the $\mathbf{e}_1$ direction. Consistent with previous study,\cite{Gong2019} we find that $C(r)$ remains finite on all cylinders even we keep up to $m=9000$ $SU(2)$ states (equivalent $m=36000$ $U(1)$ states). Surprisingly, our results show that $C(r)$ decreases notably with the increase of $L_x$ which vanishes in the long cylinder limit $L_x=\infty$ on all $L_y=6-10$ cylinders after the finite-size scaling as shown in Fig.\ref{fig:spincorJ3}(c). To test the possibility of TRS breaking in two dimensions, we have also performed the finite-size scaling of $C(r)$ as a function of $1/L_y$ by fixing the lattice ratio $L_x/L_y=3$. As an example shown in Fig.\ref{fig:spincorJ3}(d), we find that $C(L_x/4)$ decreases rapidly with the increase of $L_y$ and vanishes when $L_y$ is large enough. This indicates a possibly vanishing chiral order in the two-dimensional limit. Therefore, our results are consistent with the absence of long-range spin scalar chiral order and the QSL phase of the $J_1$-$J_2$-$J_3$ model preserves the TRS.

Our results suggest that the ground state of the $J_1$-$J_2$-$J_3$ model is consistent with a critical spin liquid with a single gapless mode. To rule out the possibility that such critical behavior could be special to the point of $J_2=0.3$ and $J_3=0.15$, we have further considered a relatively distant parameter point in the $J_1$-$J_2$-$J_3$ spin liquid phase with $J_2=0.36$ and $J_3=0.24$.\cite{Gong2019} Following the same procedure, we have observed the similar critical behavior with one gapless mode at this new point, where detailed results are provided in the SM. D and Fig. \ref{Afig:J3=0.24}. Therefore, our results suggest that the $J_1$-$J_2$-$J_3$ spin liquid is a critical phase\cite{Yao2011} instead of a critical point.

{\bf Summary and discussion: }%
We have studied the ground state properties of the spin liquid phases in both the spin-1/2 $J_1$-$J_2$ and $J_1$-$J_2$-$J_3$ models on the triangular lattice. Using large-scale parallel DMRG encoded with SU(2) spin rotational symmetry, we are able to obtain accurate results on notably longer systems by keeping a significantly large number of states in the DMRG simulation. Our results suggest that the QSL phase of the $J_1$-$J_2$ Heisenberg model is consistent with a gapped spin liquid which is characterized by the absence of gapless spin mode, short-range spin-spin and dimer-dimer correlations. In the presence of finite $J_3$ interaction, a new critical spin liquid phase emerges which has one gapless mode and quasi-long-range dimer-dimer correlation but exponentially decaying spin-spin correlation.

A striking behavior of the central charge that is prominent on cylinder geometry is that its value can be notably affected by both the boundary and finite-size effects. While long cylinders are always necessary, we find that reduced boundary effect by removing a few data points close to the open ends of the cylinders can provide more reliable results that are much closer to that in the long cylinder limit. However, it should be noted that some of the small-system behaviors, including both the central charge and various correlation functions, presented here are not special to the studies of the triangular lattice Heisenberg antiferromagnet, but also apply to various other systems as shown in previous DMRG calculations.\cite{Jiang2020,Peng2021k,Jiang2021} Our study emphasizes the perceptible effect of the finite-size and boundary effects which need to be taken into account in the numerical simulations.

%==Acknowledgements==
{\it Acknowledgments:} We would like to thank Steven Kivelson, Thomas Devereaux, Dong-Ning Sheng, Shou-Shu Gong and Hong Yao for insightful discussions. This work was supported by the Department of Energy, Office of Science, Basic Energy Sciences, Materials Sciences and Engineering Division, under Contract DE-AC02-76SF00515. Y.F.J. acknowledges the start-up grant of ShanghaiTech University. Some of the computing for this project was performed on the Sherlock cluster. We would like to thank Stanford University and the Stanford Research Computing Center for providing computational resources and support that contributed to these research results.

%reference
%

\clearpage

\begin{widetext}

\renewcommand{\theequation}{A\arabic{equation}}
\setcounter{equation}{0}
\renewcommand{\thefigure}{S\arabic{figure}}
\setcounter{figure}{0}
\renewcommand{\thetable}{A\arabic{table}}
\setcounter{table}{0}

%==appendix==
\section{Supplemental Material}
\subsection{Parallel DMRG}
In most of the DMRG simulation on the wider cylinders, a dramatically increasing bond dimension $m$ is required to capture the growing entanglement entropy of the systems, which makes obtaining converged results a challenging task. Such a heavy numerical burden can be eased by utilizing the symmetry of the Hamiltonian\cite{McCulloch2002} and parallel scheme based on distributing operators\cite{Chan2004}, where the former can reduce the numerical cost by order of magnitude and the later can speedup the computational time by $N_{node}$ times for the ideal cases. In the left panel of \Fig{Afig:pdmrg}, we sketch the parallel strategy adapted for the most time-consuming Lanczos eigen-solver. In most of DMRG studies for the condensed-matter systems, the Hamiltonian written in specific block-site-site-block decomposition contains $\sim \alpha Ly$ terms, where $\alpha$ depends on the number of the independent operators, e.g. $\{\v{S}, \v{c}^\dagger, n\}$ for the t-J model, and the complexity of the interaction. At the first iteration of the Lanczos algorithm, we equally distribute the $\alpha L_y$ terms to each node and apply them simultaneously to the initial wavefunction $\phi$ broadcasted to each node. The wave-function are then accumulated to form a new wave-function $\phi_{new}$ and broadcast to all nodes as the initial wave function of the next iteration. This procedure repeats until the convergence of Lanczos iteration is reached. Note that the communication of operators only happens for the first iteration of the Lanczos step.

Another challenge brought by the large bond dimension is the rapid growth of the local space needed for storing operators. Such issue can be eased by the parallel truncation step illustrated in the middle panel of \Fig{Afig:pdmrg}. We firstly broadcast the truncation operator $U$ to each node. The local operators of new system/environment block $O$ is distributed to all nodes and truncated by $O'=UOU^\dagger$. Instead of sending them back to the master node, the truncated operators can be directly stored in the local drive of each node. These data will be read in the similar parallel manner when we sweep back to the same position. 

One advantage of this parallel scheme is that it can produce the exactly same physical quantities as those obtained by single node DMRG since no additional approximation is introduced in the parallel steps. In the right panel of \Fig{Afig:pdmrg}, we test the speedup of the parallel algorithm by simulating the nearest neighbor t-J model on $8\times 18$ square lattice with cylinder boundary condition. The $U(1)\times U(1)$ symmetry is employed in both the single-node code and parallel code.  All the calculations are carried out using the same initial wave function and the machines with the same configuration on the Sherlock cluster at Stanford.  The comparison of the computational time costed by the Lanczos step is exhibited in the right panel of \Fig{Afig:pdmrg}, where $T_0/T_N$ is the ratio of the wall-time cost by the single node algorithm and the parallel algorithm with N nodes. In general, we observe an approximately linear scaling of $T_0/T_N$ as a function of $N_{nodes}$, with a 9-times speedup achieved by using 12 nodes. Notably, the scalability of the parallel scheme becomes better when we increase the bond dimension.

\begin{figure}[b]
    \centering
    \includegraphics[width=\linewidth]{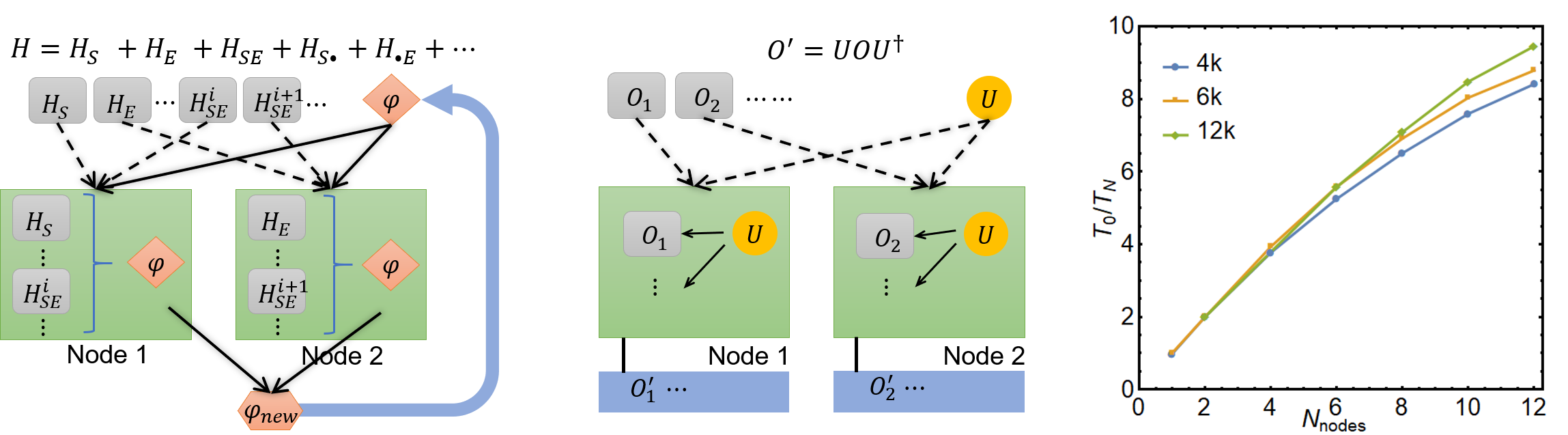}
    \caption{Left Panel: Parallel step of Lanczos algorithm. At the first step of the Lanczos iteration, each term in Hamiltonian is (approximately) equally distributed to all nodes. Here $H_{SE}^i$ denotes the $i$th operators connecting the sites in both the System and Environment part of the block-site-site-block decomposition of the Hamiltonian. After distributed Hamiltonian, the wave function $\phi$ is copied to all nodes, multiplied by the terms stored in each nodes, and accumulated to form $\phi_{new}$ used as the input wave-function of the next step of the iteration.
    Middle Panel: Parallel step of truncation and distributed storage of operators. Truncation operator $U$ is broadcast and the local operators $O$ in the enlarged basis with dimension $md$ are equally distributed to all nodes. The operator is then truncated to the new basis with dimension $m$ and stored in the local storage of each node.
    Right Panel: Wall-time of the lanczos part of parallel code with $U(1)\times U(1)$ symmetry for $t$-$J$ model on $8\times18$ square lattice with the nearest interaction. Color labels the number of kept U(1) state $m$. For a given $m$, each calculation is initialized with same wave function to obtain the exactly same output of physical quantities. We apply 4 full sweeps to get the average time.}
    \label{Afig:pdmrg}
\end{figure}

\subsection{Numerical detail}
We have applied finite truncation error extrapolation for all the physical quantities we studied to improve the accuracy of our results. A detailed example of the extrapolation is shown in Fig.\ref{Afig:scaling}(a), where the von Neumann entropy of the $J_1$-$J_2$-$J_3$ model on $L_x=30$ and $L_y=10$ cylinder is plotted as a function of truncation error $\epsilon$ of the corresponding $m=5000 \sim 9000$ number of SU(2) states. For each length of sub-cylinder $x$ (labeled by different color), we use the second order polynomial function $S(x,\epsilon)=A(x)\epsilon^2 +B(x) \epsilon +C(x)$ to extract the $S(x,\epsilon\rightarrow 0)$, or equivalently $S(x, m \rightarrow \infty)$ shown in Fig.\ref{Afig:scaling}(b). Following the same procedure, we can reliably obtain the other physical quantities at $\epsilon \rightarrow 0$ limit, e.g., the spin-spin correlation function $F(r,\epsilon \rightarrow 0)$ on the same system provided in Fig.\ref{Afig:scaling}(c). After the extrapolation, the spin-spin correlation functions slightly increase but still decay exponentially for all the cases we studied.

\subsection{Benchmark with the previous studies}
As shown in Fig.\ref{Afig:benchmark}, we benchmark our result of the $J_1$-$J_2$-$J_3$ model on the short cylinders with the previous results in Ref\cite{Gong2019}. For the $L_y=6$ cylinders, we check the entanglement entropy and central charge of the $L_x=18$ cylinders with $J_2=0.3$ and $J_3=0.15$. By keeping $m=2000$ SU(2) states, we obtain nearly same behaviour of entropy $S(x)$ and central charge $c=1.04$ as those reported in Ref\cite{Gong2019}. For the $L_y=8$ cylinder with $L_x=16$, we also find very similar $S(x)$ by keeping $m=2000\sim 4000$ SU(2) state. The extracted central charge is $c\sim 5$, similar to the one obtained in previous studies.

\begin{figure}
    \centering
    \includegraphics[width=0.88\linewidth]{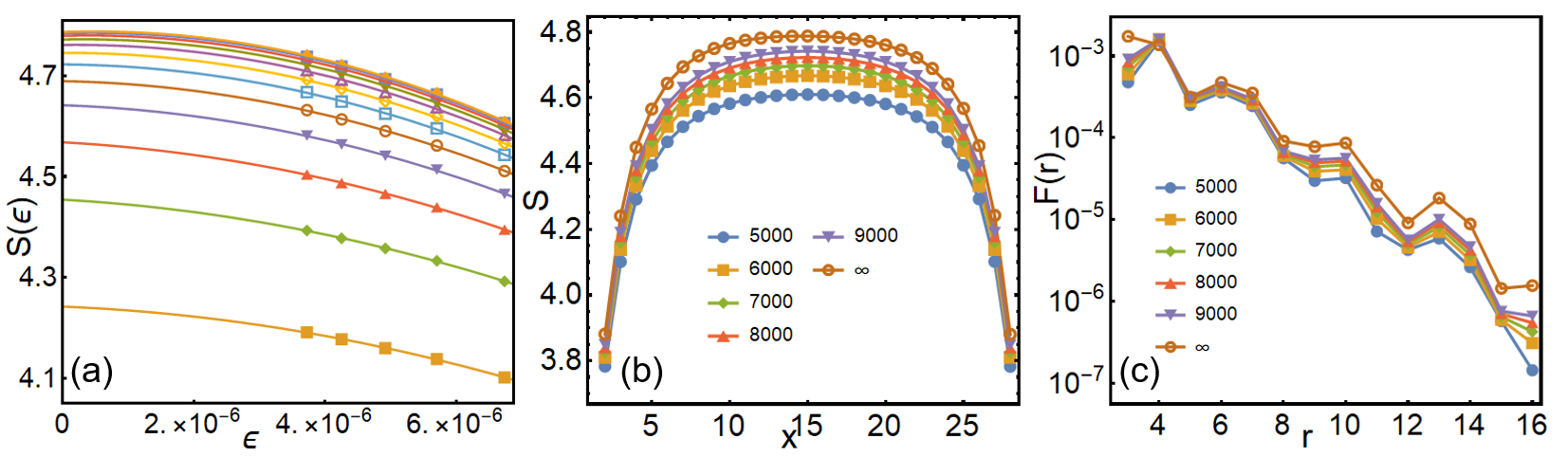}
    \caption{The finite truncation error extrapolation applied on the $J_2=0.3$ and $J_3=0.15$ model on $L_x=30$ and $L_y=10$ cylinder: (a) the von Neumman entropy $S(x, \epsilon)$ as a function of truncation error $\epsilon$, length of the sub-cylinders $x$ is labeled by different color. (b) The entropy obtained with several numbers of kept states varied from $m=5000\sim 9000$, $S(x, m\rightarrow \infty)$ is obtained from fitting shown in (a). (c) The extrapolation of the spin-spin correlation function $F(x,  m\rightarrow \infty)$.}
    \label{Afig:scaling}
\end{figure}

\begin{figure}
    \centering
    \includegraphics[width=\linewidth]{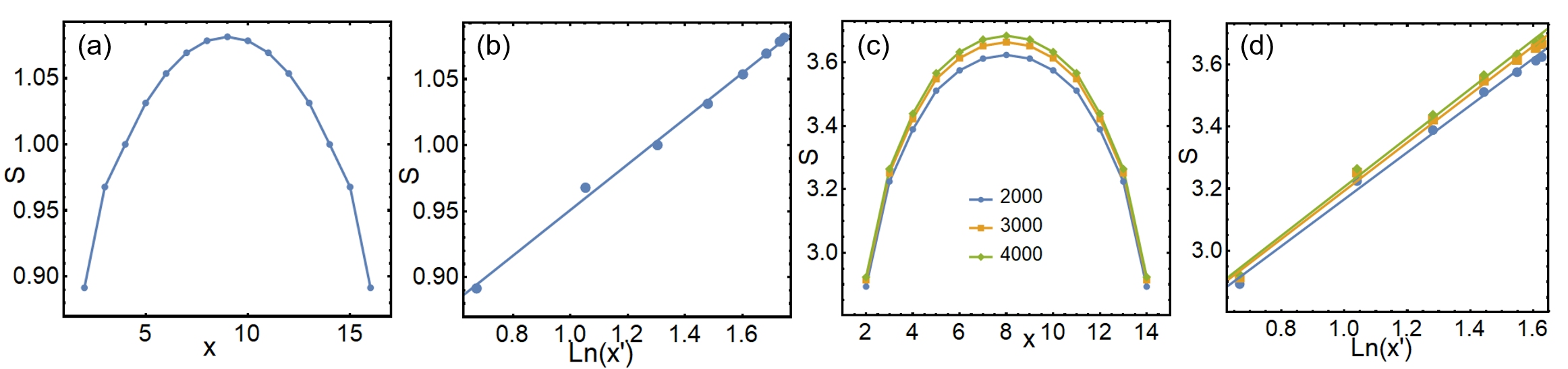}
    \caption{(a) The entanglement entropy of the $J_2=0.3$ and $J_3=0.15$ model on $18\times 6$ cylinder. (b) The central charge $c=1.04$ obtained by fitting $S(x)\sim \frac{c}{6}{\rm Ln}(x')$ where $x'=\frac{L_x}{\pi} \sin (\frac{\pi x}{L_x})$. (c) The entanglement entropy of the same model on $16\times 8$ cylinders, obtained by keeping $2000\sim 4000$ SU(2) states. (d) The central charge $c \sim 5$ extracted from the same model on $16\times 8$ cylinders.}
    \label{Afig:benchmark}
\end{figure}

\subsection{Numerical results of the $J_2=0.36$ and $J_3=0.24$ model}
For a deeper investigation of the critical spin liquid phase of the $J_1$-$J_2$-$J_3$ model, we have also considered another point with $J_2=0.36$ and $J_3=0.24$, which is fairly distant from the $J_2=0.3$ and $J_3=0.15$ point discussed in the main text. As exhibited in Fig.\ref{Afig:J3=0.24}, our results show that the ground state properties of the $J_1$-$J_2$-$J_3$ model at $J_2=0.36$ and $J_3=0.24$ are also consistent with a critical spin liquid. Following the same procedure as in the main text, we find that the central charge $c$ on the $L_y=8$ cylinder in the long $L_x$ limit is also reasonably close to $1$. The spin-spin correlation, as shown in Fig.\ref{Afig:J3=0.24}(b-c), is short-ranged with a finite correlation length $\xi_s\sim 2.24$. However, the dimer-dimer correlation $D(r)$ decays as a power-law $D(r)~r^{-K_a}$ with an exponent $K_a\sim 2.5$. To test the possibility of TRS breaking, we have also checked the scaling behavior of the spin chiral-chiral correlation $C(r=L_x/4)$ as shown in Fig.\ref{Afig:J3=0.24}(d). Similar with the case shown in the main text, we find that $C(L_x/4)$ decreases quickly with the increase of $L_x$ and drops to zero when $L_x$ is long enough. This provides more evidences on the time reversal symmetric nature of the critical spin liquid phase discussed in the main text.

\begin{figure}
    \centering
    \includegraphics[width=\linewidth]{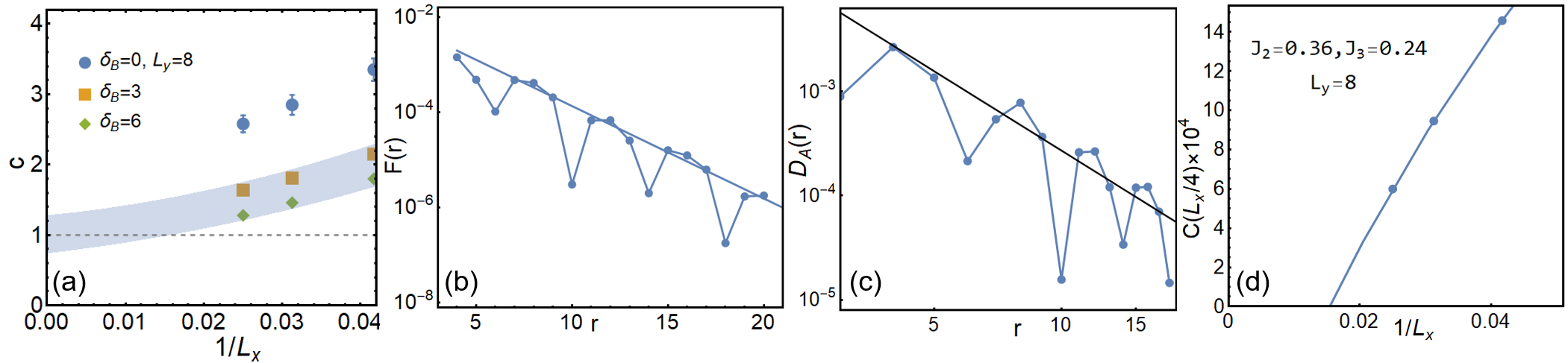}
    \caption{The numerical results of the $J_2=0.36$ and $J_3=0.24$ model on the $L_y=8$ cylinders. (a) The extract central charge $c$ as a function of $1/L_x$, where the shaded region is a guide of eyes. (b) The spin-spin correlation function $F(r)$ on the $L_x=40$ cylinder where the solid line denotes an exponential fit. (c) The dimer-dimer correlation function $D(r)$ of the $A$ bonds on the same cylinder where the solid line denotes a power-law fit. (d) The chiral-chiral correlation function $C(L_x/4)$ measured on $L_x=24-40$ cylinders. Here, we keep up to $m=10000$ SU(2) states (equivalent $m=40000$ U(1) states) in the DMRG simulation. }
    \label{Afig:J3=0.24}
\end{figure}

\begin{figure}
    \centering
    \includegraphics[width=0.8\linewidth]{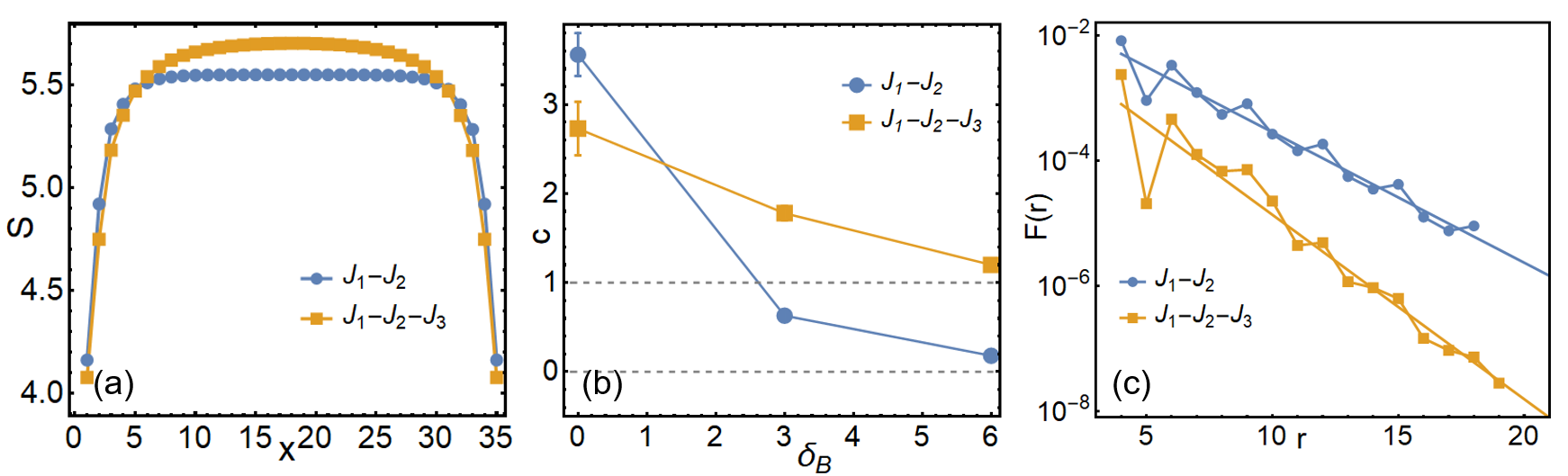}
    \caption{The preliminary result of the $J_2=0.11$ model and the $J_2=0.3$ and $J_3=0.15$ model on the $L_x=36$ and $L_y=12$ cylinders. (a) The von Neumann entropy $S(x)$ of the two models. (b) The central charge $c$ extracted from the entropy $S(x)$, we removed the $\delta_B$ point from the boundary to gradually reduce the boundary effect. (c) The exponentially decaying spin-spin correlation functions of the two systems. }
    \label{Afig:ly=12}
\end{figure}

\subsection{Preliminary results on the $L_y=12$ cylinder}
Due to the increasing entanglement entropy on the wider cylinders, reliably investigating the properties of the $L_y=12$ cylinders becomes very challenging. In this section, we discuss our preliminary result obtained on the $L_x=36$ and $L_y=12$ cylinders with $m$ up to 6000 SU(2) states. For both $J_1$-$J_2$ and $J_1$-$J_2$-$J_3$ phases, we find that the results on $L_y=12$ cylinder are qualitatively same as those on $L_y=6-10$ cylinders. In Fig.\ref{Afig:ly=12}(a) and (b), we compare the entanglement entropy and central charges of the two models. After reduced the boundary effect by gradually tuning $\delta_B$, we clearly see that the central charge of the $J_1$-$J_2$ phase monotonously approaches to $0$ while the one of the $J_1$-$J_2$-$J_3$ phase appears to saturate to the $c=1$ line, which are consistent with the properties explained in the main text. The spin-spin correlation functions are also measured on the $L_y=12$ cylinder, both of the two phases exhibit short range correlations for the spin, with the correlation length $\xi_s = 2.1(3)$ and $1.5(1)$ for the $J_1$-$J_2$ and $J_1$-$J_2$-$J_3$ phases, respectively. Remarkably, For both two phases the spin correlation length on $L_y=12$ cylinders is shorter than those measured on $L_y=8$ and $10$ cylinders, indicating that the spin gap  is likely to be finite on the 2-D system.

\end{widetext}

\end{document}